\documentclass[preprint,showpacs,preprintnumbers,amsmath,amssymb,graphicx]{revtex4}
\usepackage{amsmath,amsfonts,amssymb}
\usepackage{epsfig}
\def\bit{\begin{itemize}}
\def\eit{\end{itemize}}
\def\ben{\begin{enumerate}}
\def\een{\end{enumerate}}
\def\bed{\begin{description}}
\def\eed{\end{description}}

\def\lsim{\raise0.3ex\hbox{$<$\kern-0.75em\raise-1.1ex\hbox{$\sim$}}}
\def\gsim{\raise0.3ex\hbox{$>$\kern-0.75em\raise-1.1ex\hbox{$\sim$}}}

\let\jnfont=\rm
\def\NPB#1,{{\jnfont Nucl.\ Phys.\ B }{\bf #1},}
\def\PLB#1,{{\jnfont Phys.\ Lett.\ B }{\bf #1},}
\def\EPJC#1,{{\jnfont Eur.\ Phys.\ Jour.\ C }{\bf #1},}
\def\PRD#1,{{\jnfont Phys.\ Rev.\ D }{\bf #1},}
\def\PRL#1,{{\jnfont Phys.\ Rev.\ Lett.\ }{\bf #1},}
\def\MPLA#1,{{\jnfont Mod.\ Phys.\ Lett.\ A }{\bf #1},}
\def\JPG#1,{{\jnfont J.\ Phys.\ G}{\bf #1},}
\def\CTP#1,{{\jnfont Commun.\ Theor.\ Phys.\ }{\bf #1},}
\def\JHEP#1,{{\jnfont JHEP \ }{\bf #1},}
\def\NPPS#1,{{\jnfont Nucl.\ Phys.\ Proc.\ Suppl.\ }{\bf #1},}

\def\beq{\begin{equation}}
\def\eeq{\end{equation}}
\def\bea{\begin{eqnarray}}
\def\eea{\end{eqnarray}}
\newcommand{\ba}{\begin{array}}
\newcommand{\ea}{\end{array}}

\begin{document}

\title{Split-SUSY dark matter in light of direct detection limits}

\author{Junjie Cao$^1$,  Wenyu Wang$^2$, Jin Min Yang$^3$ }

\affiliation{
$^1$ Department of Physics, Henan Normal University, Xinxiang 453007, China\\
$^2$ Institute of Theoretical Physics, College of Applied Science,
     Beijing University of Technology, Beijing 100124, China\\
$^3$ Institute of Theoretical Physics, Academia Sinica,
              Beijing 100190, China
\vspace*{1.1cm} }

\begin{abstract}
We examine the present and future XENON limits on the neutralino
dark matter in split supersymmetry (split-SUSY). Through a scan over
the parameter space under the current constraints from collider
experiments and the WMAP measurement of the dark matter relic
density, we find that in the allowed parameter space 
a large part has been excluded by the present XENON100 limits and 
a further largish part can be covered by the future exposure (6000 kg-day). 
In case of unobservation of dark matter with such an exposure in the future, 
the lightest neutralino will remain bino-like and its annihilation is mainly through 
exchanging the SM-like Higgs boson in order to get the required relic 
density.
\end{abstract}
\maketitle

So far the only phenomenology crisis which requires new physics at
the TeV scale seems to be the cosmic dark matter. Unlike neutrino
oscillations which may indicate some new physics at a very high
unaccessible energy scale, the cosmic dark matter naturally points
to a WIMP (weakly interacting massive particle) which should appear
in some new physics around TeV scale. A perfect candidate for such a
WIMP is the lightest neutralino in low energy supersymmetry (SUSY).
As a specific low energy SUSY model, the split-SUSY \cite{split} is
phenomenologically attractive because it just gives up the acestic
(fine-tuning) problem while maintains the phenomenologically
required dark matter and the gauge coupling unification. This model
also gets rid of the notorious supersymmetric flavor problem because
of the assumed superheavy sfermions. Actually, in this framework no
scalar particles except the SM-like Higgs boson are accessible at
the foreseeable particle colliders. So the only way to explore this
model is to study its gaugino/higgsino sector, for which 
the dark matter
detection experiments like XENON \cite{xenon100} and CDMS
\cite{cdms} can interplay with the collider experiments 
to allow for a comprehensive test.

Recently, the CDMS and XENON collaborations reported their null search results
which set rather stringent limits on the dark matter scattering cross
section \cite{xenon100,cdms}. The implications of these new limits
for the neutralino dark matter in low energy SUSY models have been
discussed recently (see, e.g., \cite{Hisano,Asano,cao1}).
On the other hand, the CoGeNT \cite{CoGeNT} and DAMA/LIBRA \cite{dama}
collaborations reported some excesses which are consistent with an
explanation of a light dark matter with a mass around 10 GeV (albeit
not corroborated by CDMS or XENON results). The possible existence of
such a light dark matter also stimulated some theoretical studies in
low energy SUSY models \cite{Belanger}.

In this note we discuss the implication of the direct detection limits
for the neutralino dark matter in split-SUSY. Since the most stringent
limits come from the XENON100 results, we will focus on the present
and future (6000 kg-day) limits from XENON.
We will perform a scan over the parameter space under the current
constraints from collider experiments and the WMAP measurement of
the dark matter relic density, and display the allowed parameter
space in the plane of the dark matter scattering rate versus the dark
matter mass. Then we can see how large a parameter space can be
excluded by the present and future XENON limits.
Further, we will show the implication of XENON limits on
the properties of the neutralino dark matter and the lightest chargino.
\vspace*{0.5cm}

We start our analysis by writing out
the chargino mass matrix:
\begin{eqnarray} \label{mass1}
{{\cal M}_{\chi^\pm}}=\left( \begin{array}{cc} M_2 & \sqrt 2 m_W \sin\beta \\
                        \sqrt 2 m_W \cos\beta & \mu \end{array} \right) ,
\end{eqnarray}
where the 2-components spinors are defined as
$\tilde \psi^+=(-i\tilde \omega^+,~\tilde h^+_2)^T$,
$\tilde \psi^-=(-i\tilde \omega^-,~\tilde h^-_1)^T$.
The neutralino mass matrix is given by
\begin{widetext}
\begin{eqnarray} \label{mass2}
{{\cal M}_{\chi^0}}=\left( \begin{array}{cccc}
M_1 & 0   & - m_Z \sin\theta_W \cos\beta & m_Z \sin\theta_W \sin\beta  \\
0   & M_2 & m_Z \cos\theta_W \cos\beta   &-m_Z \cos\theta_W \sin\beta  \\
- m_Z \sin\theta_W \cos\beta &  m_Z \cos\theta_W \cos\beta  & 0 & -\mu \\
m_Z \sin\theta_W \sin\beta & -m_Z \cos\theta_W \sin\beta & -\mu & 0\\
\end{array} \right),
\end{eqnarray}
\end{widetext}
where the 2-component spinors are defined as
$\tilde \psi^0=(-i\tilde b,~-i\tilde \omega_3,
~\tilde h_1,~\tilde h_2)^T$.
In the above mass matrices, $M_1$ and $M_2$ are respectively the $U(1)$
and $SU(2)$ gaugino mass parameters, $\mu$ is the mass parameter in the
mixing term $-\mu \epsilon_{ij} H_1^iH_2^j$ in the
superpotential, and  $\tan\beta \equiv v_2/v_1$ is ratio of the
vacuum expectation values of the two Higgs doublets.

The chargino mass matrix (\ref{mass1}) is diagonalized by
$U^\ast{{\cal M}_{\chi^\pm}}V^\dagger$ to give two chargino mass
eigenstates $\chi^+_{1,2}$ with the convention
$M_{\chi^+_1}<M_{\chi^+_2}$. The eigenstates may be wino ($-i\tilde
\omega^+$) dominant or higgsino ($\tilde h^+_i$) dominant.
Similarly, the neutralino mass matrix (\ref{mass2}) is diagonalized
by $N^\ast{{\cal M}_{\chi^0}}N^\dagger$ to give four neutralino mass
eigenstates $\chi^0_{1,2,3,4}$ with the convention
$M_{\chi^0_1}<M_{\chi^0_2}<M_{\chi^0_3}<M_{\chi^0_4}$. The
neutralinos may be bino ($-i\tilde b$), wino($-i\tilde \omega_3$) or
higgsino ($\tilde h_i$) dominant. So the masses and mixings of
charginos and neutralinos are determined by four parameters: $M_1$,
$M_2$, $\mu$ and $\tan\beta$.

The spin-independent (SI) interaction between the lightest
nuetralino $\tilde{\chi}^0_1$ and the nucleon (denoted by $f_p$ for
proton and $f_n$ for neutron \cite{susy-dm-review}) is induced by
exchanging the SM-like Higgs boson or the squarks at tree level
\cite{susy-dm-review,Drees}. In split-SUSY, the squark contribution
is negligibly small, so $f_p$ is approximated by
\cite{susy-dm-review} (similarly for $f_n$)
\begin{equation}
 \begin{split}
    f_{p}  \simeq
\sum_{q=u, d, s} \frac{f_q^{H}}{m_q} m_p f_{T_q}^{(p)}
    + \frac{2}{27}f_{T_G} \sum_{q=c, b, t} \frac{f_q^{H}}{m_q} m_p,
 \end{split}     \label{2b}
\end{equation}
where $f_{Tq}^{(p)}$ denotes the fraction of $m_p$ (proton mass)
from a light quark $q$ while
$f_{T_G}=1-\sum_{u,d,s}f_{T_q}^{(p)}$ is the heavy quark
contribution through gluon exchange. $f_q^{H}$ is the
coefficient of the effective scalar operator given
by \cite{susy-dm-review}
\begin{equation}
    f_q^{H} = m_q \frac{g_2^2}{4 m_W}
     \frac{C_{h \tilde \chi \tilde \chi}  C_{hqq}}{m_{h}^2}.
    \label{Higgs-contr}
\end{equation}
with $C$ standing for the corresponding Yukawa couplings. The
$\tilde \chi^0$-nucleus scattering rate is then given by
\cite{susy-dm-review}
\begin{equation}
    \sigma^{SI} = \frac{4}{\pi}
    \left( \frac{m_{\tilde \chi^0} m_T}{m_{\tilde \chi^0} + m_T} \right)^2
    \times \bigl( n_p f_p + n_n f_n \bigr)^2,
\end{equation}
where $m_T$ is the mass of target nucleus and $n_p (n_n)$ is the number of
proton (neutron) in the target nucleus.

From the above formulas we can infer in which situation the
scattering cross section is large. Eq.(\ref{Higgs-contr}) indicates
that this occurs when $C_{h \tilde \chi \tilde \chi}$ and/or
$C_{hqq}$ get enhanced.  As the Higgs boson is SM-like, $C_{hqq}$
has no $\tan \beta$ enhancement for the down-type quark. We here
only check the behavior of $C_{h \tilde \chi \tilde \chi}$ with the
variation of the relevant SUSY parameters. For a bino-like
$\tilde{\chi}^0_1$, this coupling is generated through the
bino-higgsino mixing and thus a large $C_{h \tilde \chi \tilde
\chi}$ needs a large mixing, which means a small $\mu$. To make this
statement clearer, we consider the limit $M_1 \ll M_2, \mu$ ($M_1$,
$M_2$ and $\mu$ denoting respectively the mass of bino, wino and
higgsino). After diagonalizing the neutralino mass matrix in a
perturbative way, one can get \cite{Hisano}
\begin{equation}
 \begin{split}
    &C_{h \tilde \chi \tilde \chi} \simeq
  \frac{m_Z \sin \theta_W \tan \theta_W}{M_1^2 - \mu^2}
    \bigl[ M_1 + \mu \sin2 \beta \bigr]. \\
 \end{split}
\label{2d}
\end{equation}
So the coupling $C_{h \tilde \chi \tilde \chi}$ becomes large when
$\mu$ approaches downward to $M_1$.

In our numerical calculation for the dark matter-nucleon scattering rate,
we considered all the contributions (including QCD corrections) known so far.
We take $f_{T_u}^{(p)} =0.023$, $f_{T_d}^{(p)} = 0.032$, $f_{T_u}^{(n)} = 0.017$,
$f_{T_d}^{(n)} = 0.041$ and $f_{T_s}^{(p)} = f_{T_s}^{(n)} = 0.020$
\cite{Djoudi,Carena,Hisano:2010ct}.
Note that here the value of $f_{T_s}$ is much smaller than that
taken in most previous studies. This small value comes from the
recent lattice simulation \cite{lattice}, and it can reduce the
scattering rate significantly.

For the calculation of the SM-like Higgs boson mass, since 
in split-SUSY we have $\log(m_{\tilde{f}}^2/m_t^2) \gg 1$ which
will spoil the convergence of the traditional loop expansion in
evaluating the SUSY effects on the Higgs boson self-energy, 
so we use the effective potential method which involves the
renormalization group evolution of the SUSY effects from the squark
scale to the electroweak scale \cite{Binger:2004nn}. This computation
method is employed in the package NMSSMTools \cite{nmtool}. 
This package, which primarily 
acts as an important tool for the study of the phenomenology of the 
Next-to-Minimal Supersymmetric Model, can also be applied 
to the MSSM case by setting $\lambda = \kappa$ approach zero
(with this setting the singlet superfield decouples from the rest of
the theory so that the MSSM phenomenology is recovered \cite{Ulrich}). 
Throughout our calculations we use this package.

As shown in \cite{Wang:2005kf}, the effects of the sfermion and the
heavy Higgs bosons on electroweak theory begin to decouple when the
particles are heavier than several TeV. So in our analysis we set
$m_{\tilde{f}} = m_A \equiv M_0 = 10 {\rm TeV}$ and the trilinear term
$A_t=A_b=0 {\rm TeV}$ to simulate the split-SUSY scenario. 
We checked that the results with $M_0 = 100 {\rm
TeV}$ are quite similar to the results with $M_0 = 10 {\rm TeV}$. 
We also checked that the package DarkSUSY \cite{darksusy}, which
calculates the Higgs mass by loop expansion, can yield 
similar results if we set $M_0 = 10{\rm TeV}$, but it
does not work if we choose $M_0 = 100{\rm TeV}$.

The remained SUSY parameters are $\tan\beta$, $M_1$, $M_2$, $M_3$,
$\mu$. We assume the SUSY GUT relation for the gaugino
masses, i.e. $M_1=(5s_W^2/3(1-s_W^2))M_2\simeq 0.5M_2$ and
$M_3=(\alpha_s s_W^2/\alpha_{\rm EW}) M_2\simeq 3 M_2$ at the
electroweak scale, and thus we only have three parameters to 
explore. In our analysis, we scan these parameter in the ranges
\begin{eqnarray}
1<\tan\beta<50, ~~0{\rm ~GeV}<M_2,~\mu< 800{\rm ~GeV}.
\end{eqnarray}
In our scan we consider the following constraints:
(1) $\tilde{\chi}_1^0$ to account for the WMAP measured dark matter
    relic density at $2\sigma$ level \cite{wmap};
(2) The LEP lower bounds on the Higgs boson,
    neutralinos and charginos, including the $Z$-boson
    invisible decay;
(3) The precision EW observables plus $R_b$ \cite{Cao}. The samples
surviving the above constraints will be input for the calculation of
the $\tilde{\chi}-N$ scattering rate. 

\begin{figure}[htbp]
\includegraphics[width=15cm]{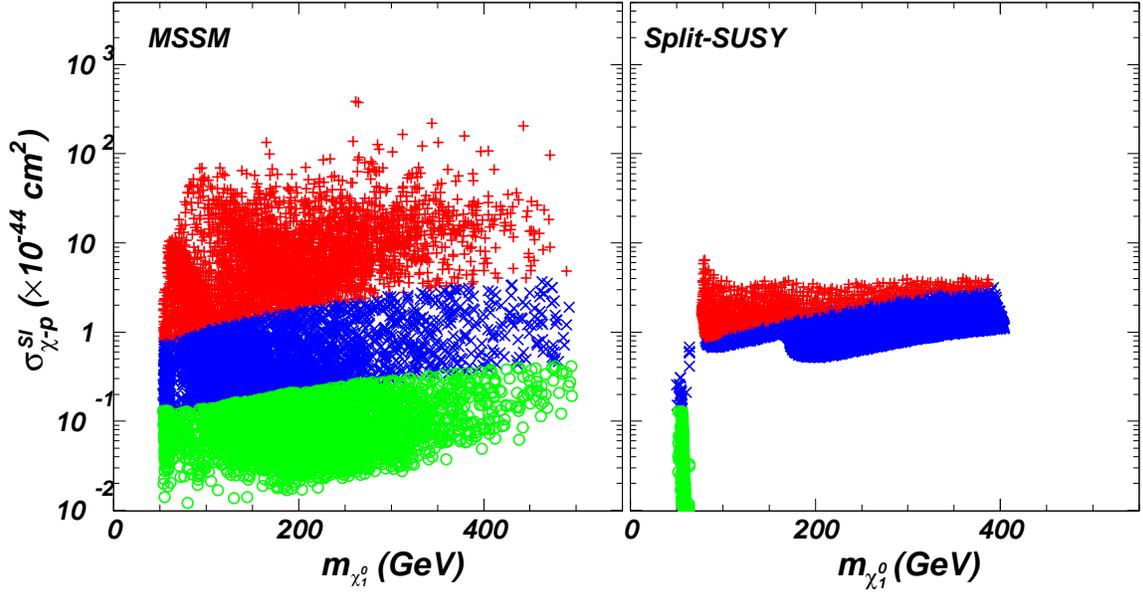}
\vspace{-0.7cm} \caption{The right (left) panel is the scatter
plots of the split-SUSY (MSSM) parameter space which survived the
constraints from the dark matter relic density ($2\sigma$) and the
collider experiments. The `$+$' points (red) are excluded by
XENON100 (90\%\ C.L.) limits, the `$\times$' (blue) will be
covered by the future XENON exposure (6000 kg-days), and the
`$\circ$' (green) are beyond the future XENON sensitivity.}
\label{fig1}
\end{figure}
Our scan samples are $10^{7}$ random points in the parameter space,
and about 14000 samples can survive the constraints from the dark
matter relic density ($2\sigma$) and the collider experiments. The
survived samples are displayed in Fig.~\ref{fig1} in comparison with
the MSSM results taken from \cite{cao1}. From the figure we can see
that although the $\tilde{\chi}-N$ scattering cross section is
highly suppressed in split-SUSY, still lots of samples 
can be excluded by the present XENON100 (90\%\ C.L.)
limits, and the future exposure of XENON (6000 kg-days) can further 
cover a large part of the survived parameter space.  
In case of null results in the future XENON experiment, 
the remained parameter space is characterized by  
$ 50{\rm GeV} < m_{\tilde\chi^0_1} < 75 {\rm GeV}$ with 
$\tilde{\chi}^0_1$ mainly annihilating
through exchanging the SM-like Higgs boson to
get the measured relic density \cite{Feldman}.

\begin{figure}[htbp]
\scalebox{0.5}{\epsfig{file=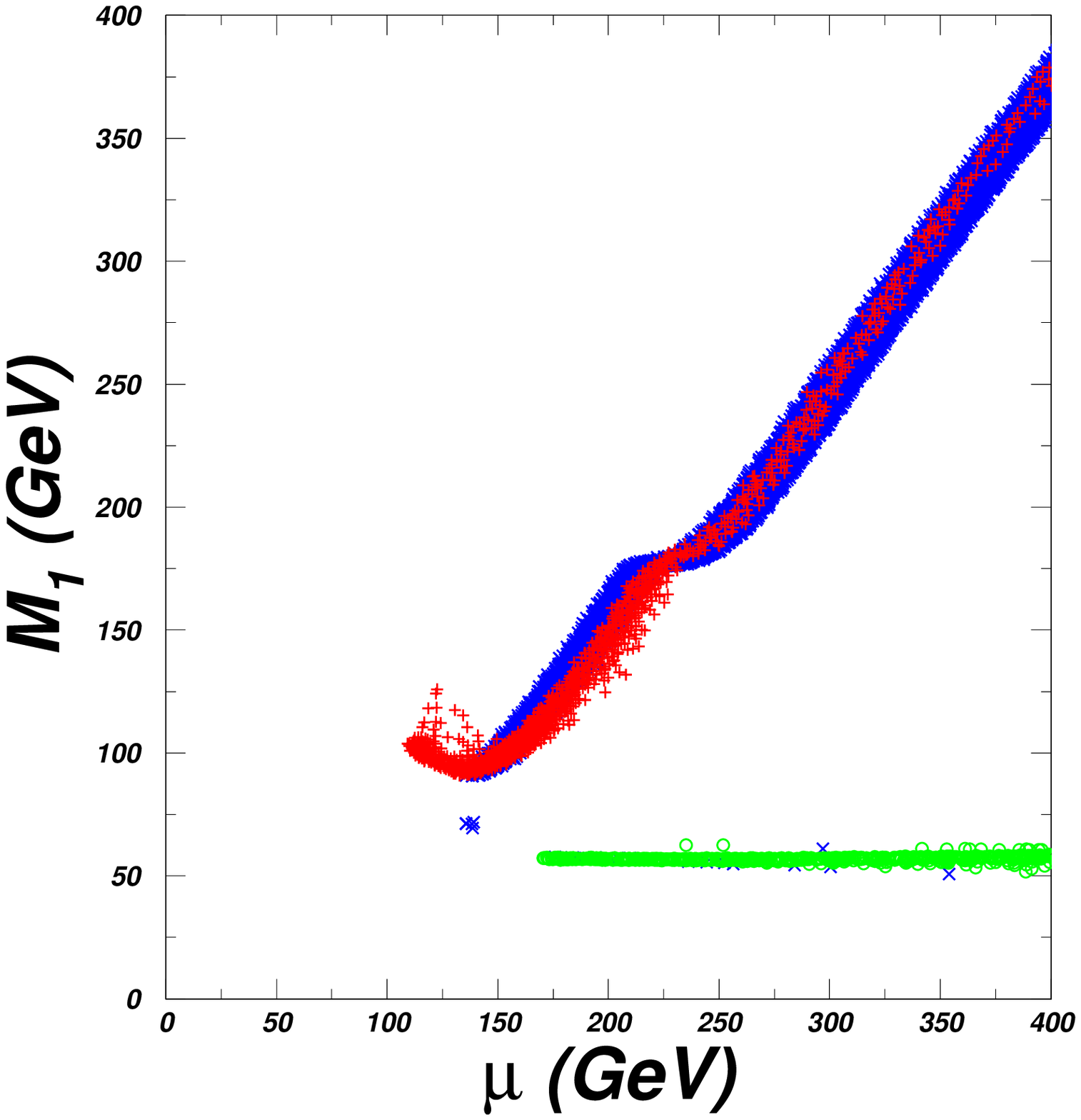}}
\scalebox{0.5}{\epsfig{file=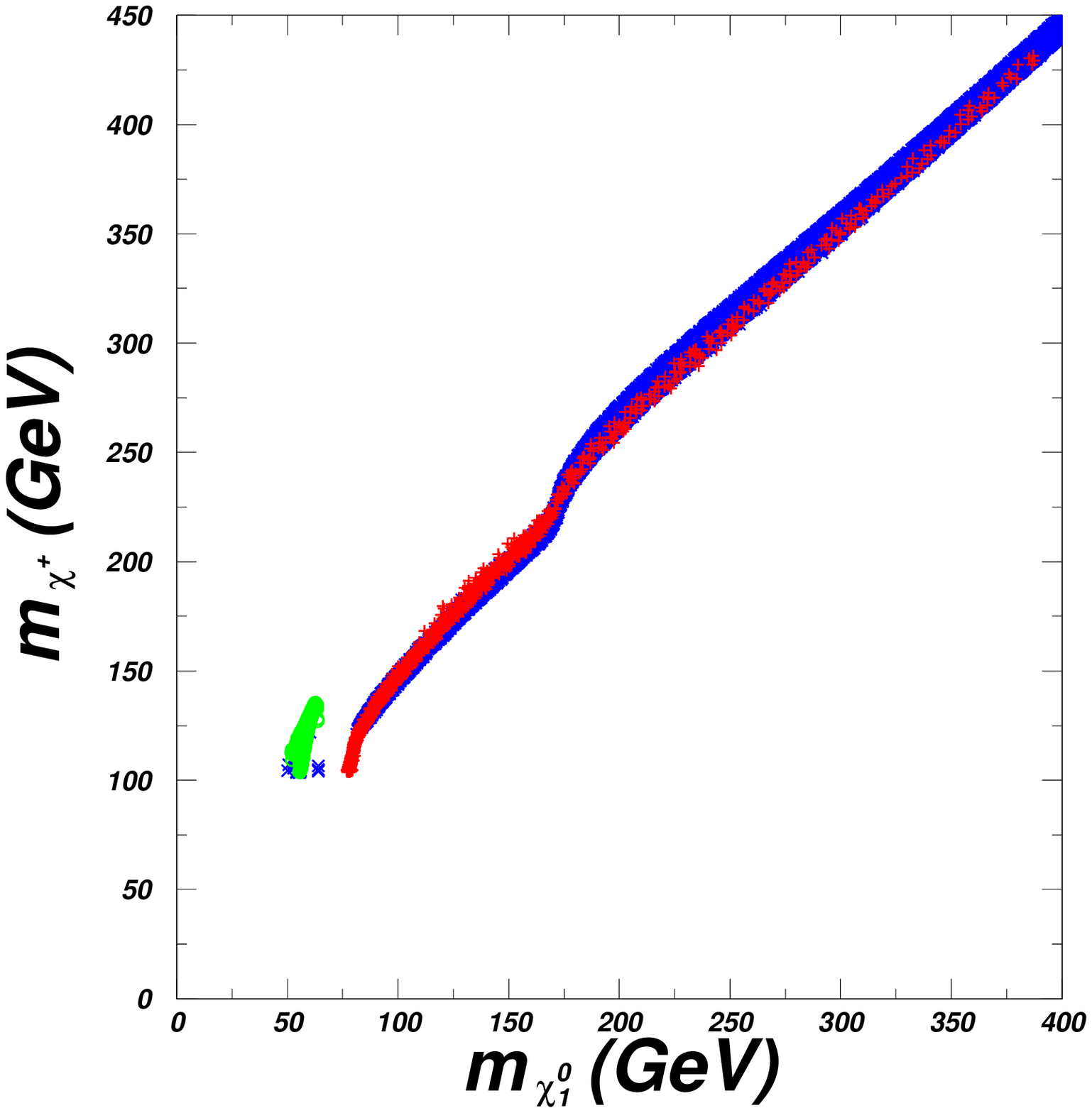}}
\vspace{-0.7cm}
\caption{Same as the right panel of Fig.~\ref{fig1}, but showing $M_1$
 versus $\mu$ and $m_{\tilde \chi^+_1}$ versus the dark matter mass.}
\label{fig2}
\end{figure}
In the following we check the properties of the parameter space
surviving the present XENON experiment. As shown in Eq.(\ref{2d}),
as $\mu$ approaches downward to $M_1$, the coupling
$C_{h\tilde\chi\tilde\chi}$ can be enhanced. This is reflected in
the left panel of Fig.\ref{fig2}, where one can learn that, for most
points excluded by the present XENON limits or covered by the future
XENON exposure, they are in the region of $M_1\simeq \mu$ so that
the $\tilde{\chi}-N$ scattering cross section is large.  In
contrast, the remained unaccessible points go into a region 
(denoted by green `$\circ$') in which $\mu$ is much larger than $M_1$. 
This unaccessible region is shown again in the right panel of Fig.~\ref{fig2} 
which indicates that $m_{\chi^+_1}$ is about $2m_{\chi_1^0}$. The reason is 
in this  region the lightest neutralino is bino-like and the lightest 
chargino is wino-like.

\begin{figure}[htbp]
\scalebox{0.75}{\epsfig{file=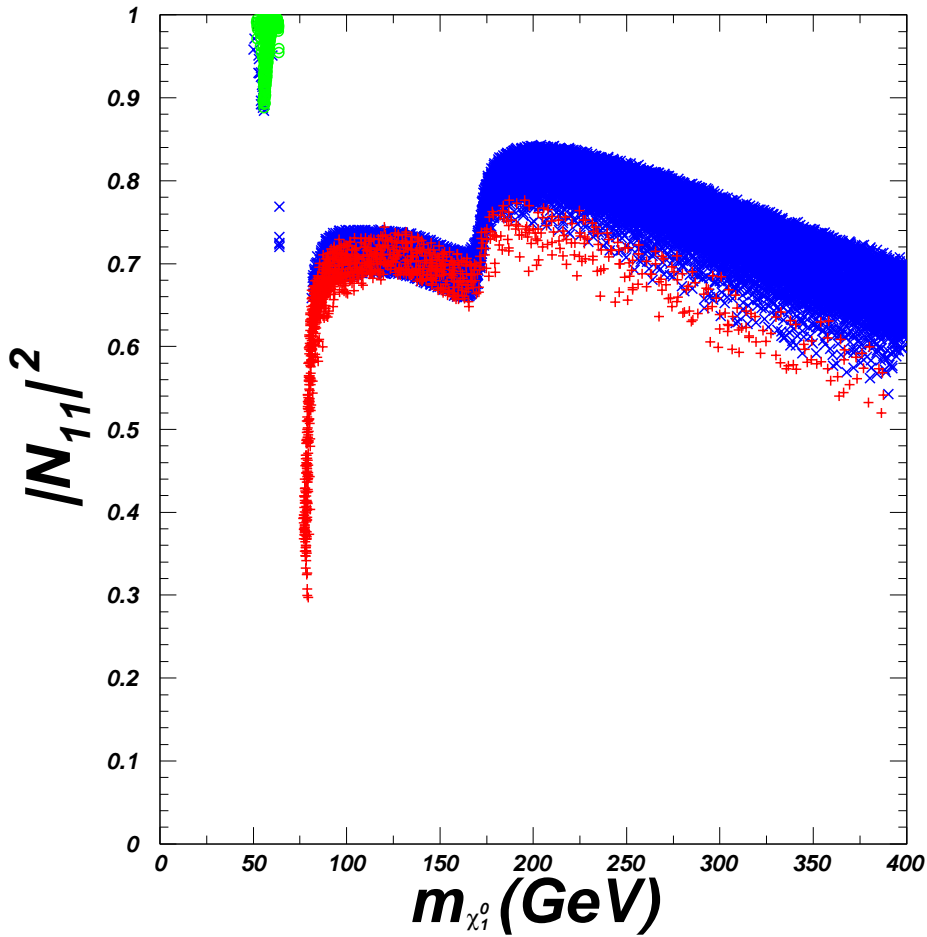}}
\vspace{-1.0cm}
\caption{Same as the right panel of
Fig.~\ref{fig1}, but showing the bino component
of the lightest neutralino.}
\label{fig3}
\end{figure}
\begin{figure}[htbp]
\scalebox{0.5}{\epsfig{file=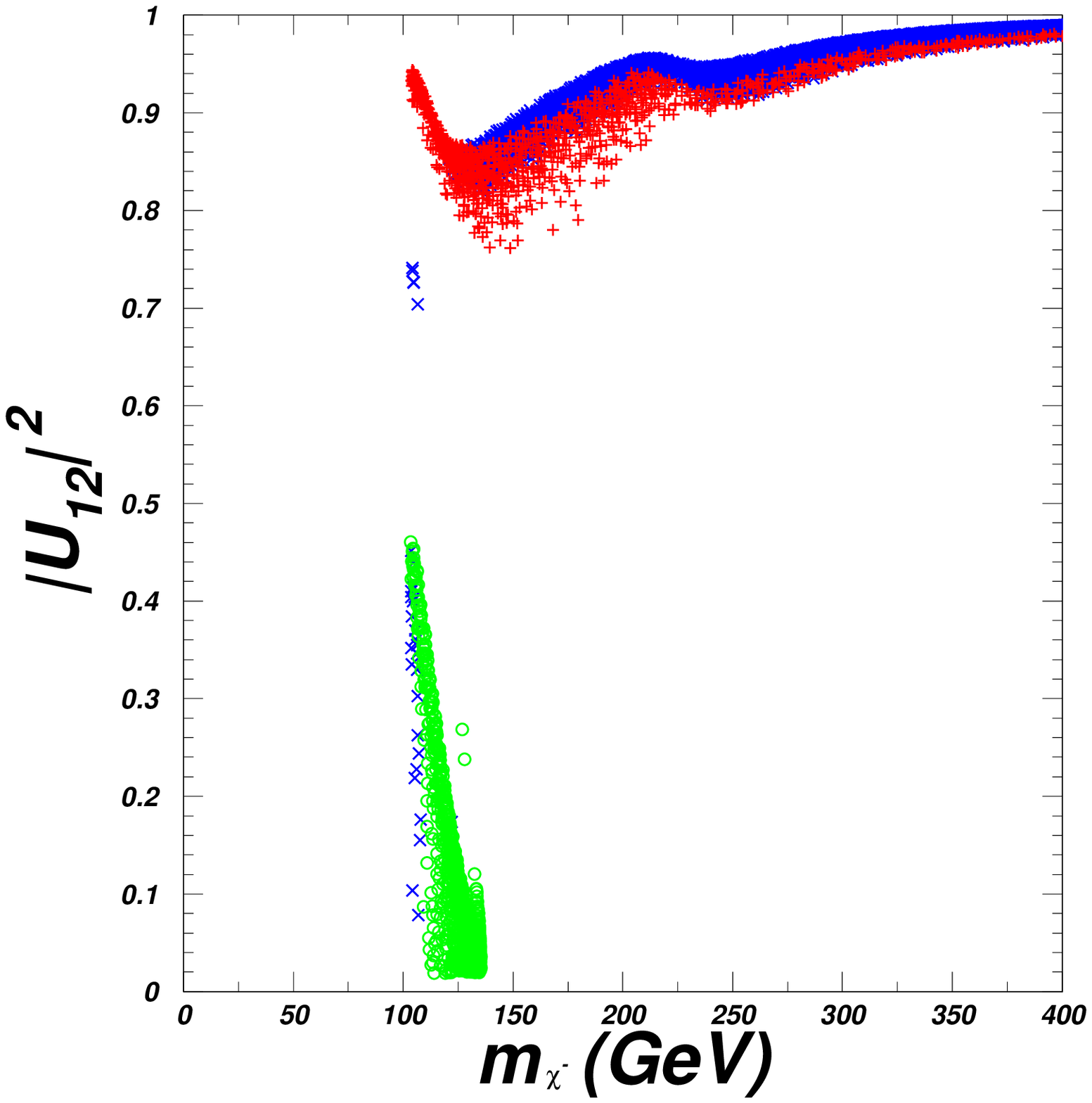}}
\scalebox{0.5}{\epsfig{file=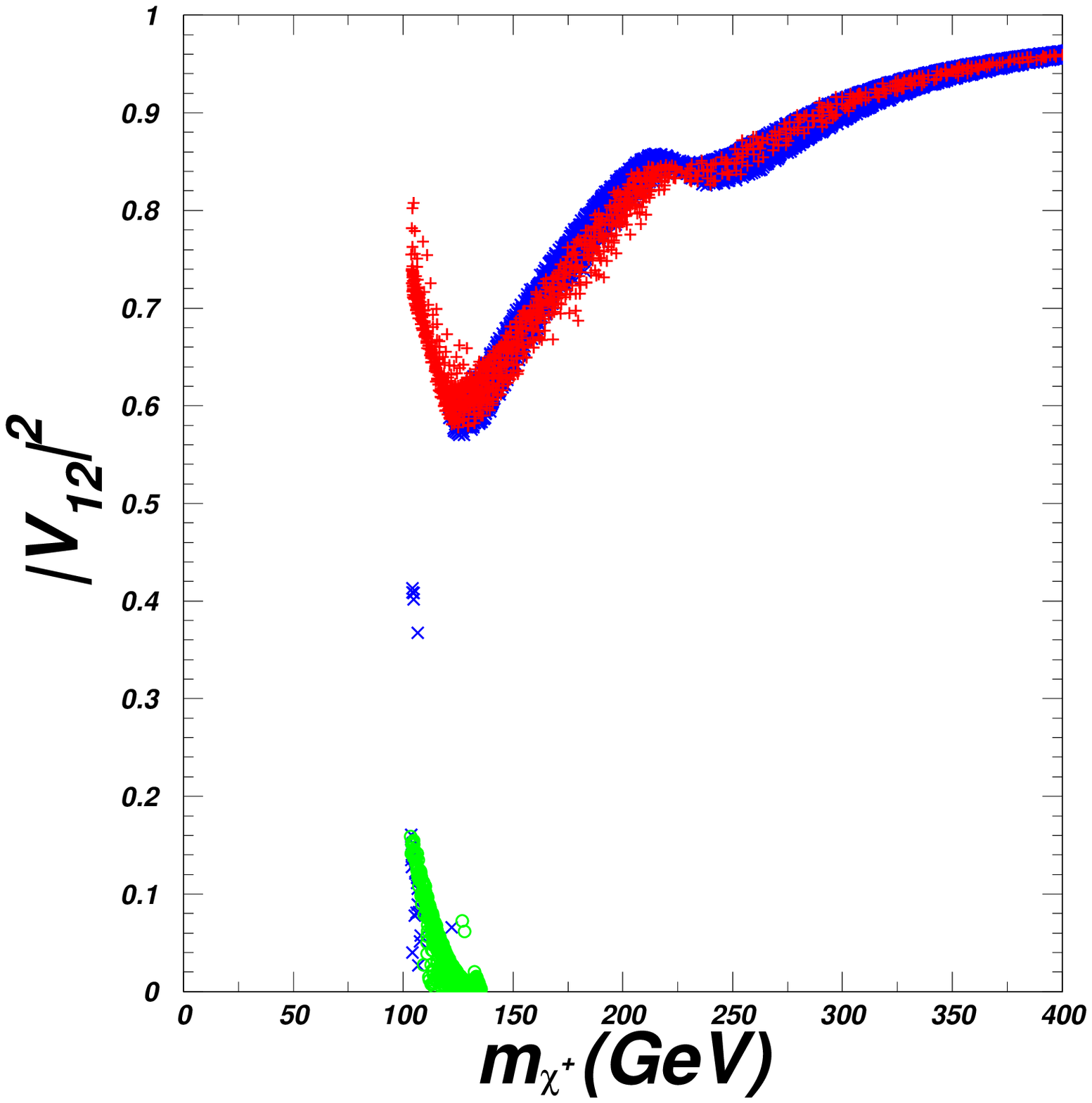}}
\vspace{-0.8cm}
\caption{Same as the right panel of
Fig.~\ref{fig1}, but showing the higgsino
component of the lightest chargino.}
\label{fig4}
\end{figure}

We show the bino component of the lightest neutralino
in Fig.~\ref{fig3} and the higgsino component of the lightest chargino
in Fig.~\ref{fig4}. From Fig.~\ref{fig3} we see that the neutralino
is accessible when its bino component is small
(higgsino component is large), while in the unaccessible region 
the neutralino is highly bino-like.
From Fig.~\ref{fig4} we see that in the  accessible region
the chargino has a large higgsino
component (a small wino component), while in the unaccessible region 
the chargino has a small higgsino component (a large wino component).

\begin{figure}[htbp]
\hspace{-0.5cm}\vspace{-0.5cm}
\scalebox{0.8}{\epsfig{file=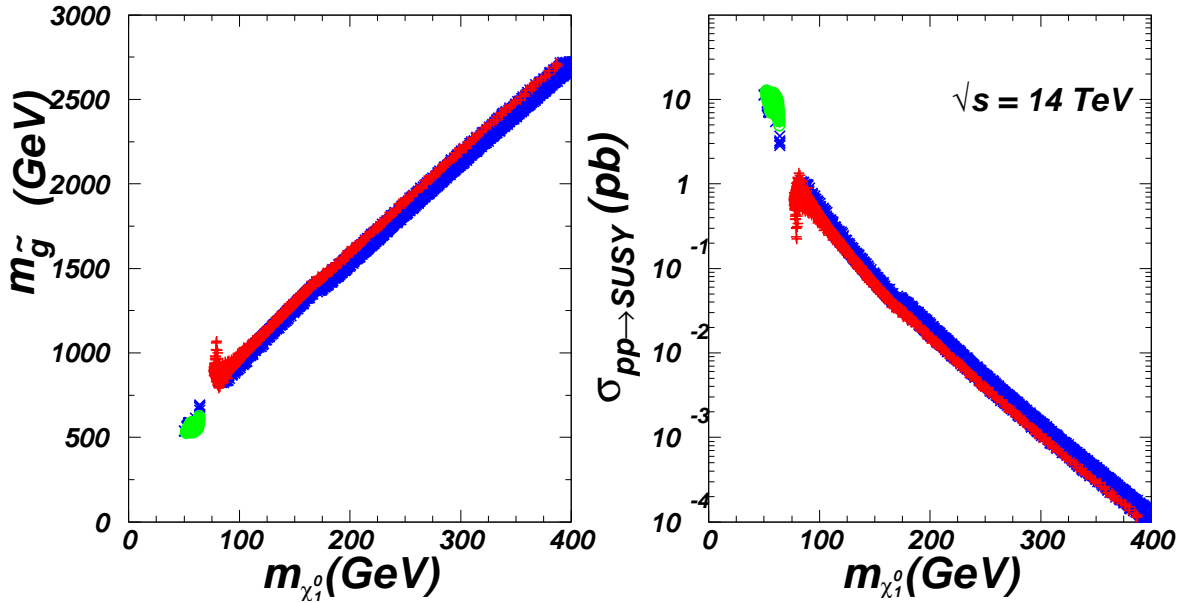}}
\caption{Same as the right panel of
Fig.~\ref{fig1}, but showing the gluino mass
and the gluino pair production cross section at the LHC.}
\label{fig5}
\end{figure}
Since the lightest neutralino is bino-like except in the $M_1
\simeq \mu$ region, the mass of gluino has an approximate 
linear relation with $m_{\tilde \chi^0}$ which is shown in left panel of Fig.
\ref{fig5}. The peak in the plot is the region where the lightest
neutralino is higgsino dominant. The dominant production of
split-SUSY particles at the LHC is $pp\to \tilde g \tilde g$, whose
cross section is shown in the right panel of Fig. \ref{fig5}. We can
see that the region unaccesible at XENON has a large production rate at
the LHC. The observability of such a light gluino pair production 
has been studied in \cite{last}, which found $S/\sqrt{B}$ can reach 
23 from ATLAS 0-lepton search at the LHC-7 with 5 fb$^{-1}$ integrated 
luminosity. So we conclude that the LHC and XENON will play 
complementary roles in testing split-SUSY.

Note that in split-SUSY the neutralino dark matter cannot be as light as
several GeV to explain the CoGeNT and DAMA/LIBRA results. As shown in
\cite{Belanger}, the neutralino dark matter in the MSSM cannot be such
light either; only in the framework of the Next-to-Minimal Supersymmetric 
Model can the dark matter be so light and have a large scattering cross
section with the nucleon to explain the CoGeNT and DAMA/LIBRA results.
\vspace*{0.5cm}

In conclusion, we studied the present and future XENON limits on the
neutralino dark matter in split-SUSY.
We performed a scan over the parameter space under the current constraints
from collider experiments and the WMAP measurement of the dark matter
relic density. We found that in the allowed parameter space a large
part has already been excluded by the present XENON100 limits while
a further largish part can be covered by the future exposure (6000 kg-days).
In case of unobservation of dark matter in the future exposure of XENON, 
the lightest neutralino will be constrained to be bino-like and the 
lightest chargino will be a light wino-like one below 150 GeV.
\vspace*{0.5cm}

JMY thanks JSPS for the invitation
fellowship (S-11028) and the particle physics group of
Tohoku University for their hospitality.
This work was supported in part by NSFC
(Nos.~10821504, 10725526, 10635030, 11075045, 11005006) and the
Doctor Fundation of BJUT (No. X0006015201102).

\end{document}